# A hybrid deep learning approach for purchasing strategy of carbon emission rights -- Based on Shanghai pilot market

Jiayue Xu, Yi Fu

School of Finance and Business, Shanghai Normal University, Shanghai 200234, China

Correspondence should be addressed to Yi Fu; fuyi@shnu.edu.cn

**Abstract:** The price of carbon emission rights play a crucial role in carbon trading markets. Therefore, accurate prediction of the price is critical. Taking the Shanghai pilot market as an example, this paper attempted to design a carbon emission purchasing strategy for enterprises, and establish a carbon emission price prediction model to help them reduce the purchasing cost. To make predictions more precise, we built a hybrid deep learning model by embedding Generalized Autoregressive Conditional Heteroskedastic (GARCH) into the Gate Recurrent Unit (GRU) model, and compared the performance with those of other models. Then, based on the Iceberg Order Theory and the predicted price, we proposed the purchasing strategy of carbon emission rights. As a result, the prediction errors of the GARCH-GRU model with a 5-day sliding time window were the minimum values of all six models. And in the simulation, the purchasing strategy based on the GARCH-GRU model was executed with the least cost as well. The carbon emission purchasing strategy constructed by the hybrid deep learning method can accurately send out timing signals, and help enterprises reduce the purchasing cost of carbon emission permits.

# 1 Introduction

In recent years, with the development of the global economy, carbon emissions have increased day by day, and the resulting environmental pollution problem has become one of the crucial challenges today. To reduce greenhouse gas emissions, the greenhouse gas content in the atmosphere should be stabilized at an appropriate level to prevent drastic climate change from damaging the environment. With the introduction of *the United Nations Framework Convention on Climate Change* (1992) and *Kyoto Protocol* (1997), carbon finance came into being. In 2016, more than 170 countries signed the *Paris Agreement*. It clearly stated that the world must rapidly reduce carbon emissions to 25 billion tons by 2030. With only nine years left, all countries must accelerate the implementation of their emission reduction responsibilities. As a developing country, China has played a leading role in the global carbon emission reduction initiative.

At present, there are two types of market ways to control greenhouse gas emissions: carbon tax and carbon market. The former is a price-based tool, and the latter is a quantity-based tool. As of April 2020, there are 31 carbon emission trading markets and 30 carbon tax mechanisms worldwide. China is also actively developing the carbon trading market, and the main trading product is carbon emission rights. The annual carbon emission rights allocated to the company are free, and the amount is usually determined based on historical data. However, according to the performance of specific companies, their annual carbon allowances will be surplus or insufficient, which creates a demand for carbon emission rights trading. For companies, carbon trading will undoubtedly increase the cost, and it is a compulsory measure to urge emission reduction. For the countries, this market-based revenue can partially replace policy subsidies and promote renewable energy to gradually replace fossils energy, thereby driving the transformation of energy structure and ultimately achieving carbon neutrality.

In 2011, Beijing, Shanghai, Shenzhen, and other seven areas launched local carbon trading pilot projects, covering about 3000 companies in more than 20 industries such as steel and power. The cumulative transaction volume of seven pilot markets is more than 400 million tons, and the cumulative transaction volume of which is more than 9 billion yuan. In 2020, the pilot carbon markets were affected by epidemic factors, and the transaction volume has decreased compared with that of 2019. However, the average transaction price has risen sharply, reaching 27.48 yuan per ton, 23% higher than the 22.30 yuan per ton in 2019. It means that companies must formulate carbon emission rights trading strategies to save costs scientifically. Therefore, this paper used a hybrid deep-learning method, taking the Shanghai carbon trading market as an example, to dynamically predict the price of carbon emission rights. On this basis, companies need to formulate a scientific and reasonable purchasing strategy.

The follow-up of the paper is divided into four parts. First of all, this paper reviewed the previous research on the price of carbon emission rights and proposed this paper's marginal contribution. Secondly, this paper analyzed the influencing factors of carbon emission rights' price and determine the input variables required for the price prediction of carbon emission rights. Then we used GARCH and recurrent neural networks to establish a price prediction model for carbon emission rights and designed carbon emission rights purchasing strategies. Finally, combined with the research content and calculation results, we achieved research conclusions. Our main contribution is to establish a hybrid deep learning model with a rolling frame based on Kim's [19] and Yan's studies [20], and we apply the model to the price prediction of the carbon trading market. It is proved that the prediction performance of this method is better than that of some other traditional models. The strategy based on the prediction can help enterprises to reduce costs. Furthermore, the importance of GARCH prediction in the variable set is ranked by the recursive feature elimination method.

## 2 Literature Review

The preliminary research related to this paper is divided into the following three categories:

One type of research focused on the analysis of factors affecting the price of carbon emission rights. Chen[1] took the EU emissions trading system as the research object and conducted theoretical analysis from four aspects: supply, demand, market, and weather. The supply of allowances affected by the policy system is the most critical factor affecting the transaction price. The improvement of the trading system had gradually weakened its influence. The study also found that energy prices such as crude oil, natural gas, and coal were also the main factors affecting carbon emission rights. Among them, the price of coal had the most significant impact, while weather factors such as wind speed, temperature, and precipitation had little effect on the price of carbon emission rights. Later, Zou[2] proposed in the study that the spot price of carbon emission certified emission reductions (CERs) was positively affected by macro-economic indicators (industrial production index) and climate indicators, among environmental factors. In addition, the price of carbon asset futures was also one of its essential influencing factors. Different from the former viewpoint, Ma et al.[3] analyzed the carbon emission rights price in the Beijing pilot market and believed that the price was negatively correlated with the price of international CERs and the level of industrial development. At the same time, the study assumed that the price of carbon emission rights was positively correlated with the price of traditional energy and the prosperity of the financial market. Considering that the domestic market would be in line with the international market after its development gradually matures, Zhou et al.[4] also considered the CERs futures price factor in the EU ETS market based on these studies. Li[5] further constructed a VAR model on the factors affecting the price of carbon emission rights. The model results showed that the price of natural gas, crude oil, and macroeconomics in the price of primary energy had a one-way positive impact on the pricing of carbon emission rights trading, and coal price was a one-way negative effect on carbon emission trading pricing. In addition, air quality and carbon emission prices influence each other. The results also showed that there was no influencing relationship between abnormal weather and carbon emission trading pricing.

Another type of research focused on the prediction of financial time series. Standard financial time series forecasting methods included traditional measurement economics methods, machine learning methods, and new research methods which mixed the two methods. In 1982, Engle[6] found in his research that the probability distribution of financial time series data generally had a spike and fat tail shape. This study proposed an Autoregressive Conditional Heteroskedasticity (ARCH) model, which could reflect volatility agglomeration and fat-tail phenomenon. Bollerslev[7] constructed the conditional variance based on the ARCH with the autoregressive moving average model (Autoregressive Moving Average Model, ARMA) morphology and established the GARCH model. Moreover, the exponential GARCH model (EGARCH)[8], asymmetric power GARCH model (APGARCH)[9], ARCH model with institutional transformation (SWARCH)[10], were proposed in the past 20 years. These GARCH family models have been widely used in fitting financial time series. For example, Zhang[11] took the four A-share market indexes as the empirical research objects and established asymmetric GARCH family models. In this study, the Power ARCH model (PARCH) could capture the leverage effect of volatility, and further reduced the information criterion value of the model. Zhang et al.[12] studied the Shenzhen carbon trading market. The GARCH model was used to describe the price fluctuation and risk formation mechanism of carbon emission rights. They believed that the domestic carbon emission price stabilized after 2013, but after 2018, affected by economic downturn and market information asymmetry, domestic carbon emission price fluctuations began to increase.

In recent years, with the rise of machine learning, a class of machine learning models called Recurrent Neural Networks (RNNs) has begun to enter the field of vision of researchers. Unlike traditional neural networks, RNNs introduce a directional loop path in the model, which can deal with sequence correlation between inputs. The most widely used RNNs model is the Long Short Term Memory Network (LSTM)[13]. This type of model can well fit the nonlinearity and volatility agglomeration characteristics of time series. For example, Fister.D [14] used LSTM to construct a machine learning system for automatic stock trading. The results showed that the strategy performed better than other conventional trading strategies. Compared with traditional econometric methods, machine learning models have certain advantages in fitting the nonlinear characteristics of time series. Therefore, many literary studies have shown that they performed well in predicting financial time series. For example, Zheng[15] used the BP artificial neural network model to identify the risk of Shenzhen's carbon trading price. Li[16] established a two-layer LSTM with ten technical indicators and the rise and fall of the closing price as a categorical output variable. As a result, the performance of the two-layer LSTM was better than the traditional linear fitting model, as well as support vector machines, decision trees, so this type of model had received widespread attention.

In the third aspect, scholars have realized that the first two methods have their advantages, and gradually merged the two models. Hajizadeh et al.[17] proposed a model that combines Artificial Neural Network (ANN) with GARCH. The study showed that a hybrid ANN and GARCH model usually had more advantages than a single ANN or time series model. Similarly, Kristjanpoller et al.[18] took Brazil, Chile, and Mexico as examples of the three Latin American stock exchange indexes, and used the hybrid artificial neural network model ANN-GARCH, which significantly improved the prediction performance of the single GARCH model in the Latin American market. Furthermore, Kim[19]; Yan et al.[20] tried to integrate the values and parameters of the GARCH family model into the LSTM model, and respectively predicted the volatility of the stock price and the volatility of the copper transaction price. The empirical results showed that combining GARCH forecasts with domestic and foreign market factors could well fit linear information characteristics to the time series, and significantly improved the predictive ability of the neural network model. Xu et al. [21] proposed a time-series data prediction hybrid model combining Linear Regression (LR) model and Deep Belief Network (DBN) model. In this hybrid model, different single models were used to capture linear and nonlinear behavior of time series respectively. In Zhang's research [22], they proposed a new hybrid deep learning model, which utilized the advantages of the time series decomposition technique: Empirical Mode Decomposition (EMD) and a hyperparameter optimization algorithm Tree of Parzen Estimator (TPEs) to predict sugar prices.

In summary, the preliminary research related to this paper mainly focused on analyzing the influencing factors of carbon emission prices and applying machine learning methods in traditional financial asset price forecasting. Based on Kim's [19] and Yan's studies[20], the marginal contribution of this paper includes two aspects. Firstly, this paper used a hybrid deep learning method to predict carbon emission rights price and builds a price prediction model for the Shanghai pilot market. This study proposed carbon emission rights purchasing strategies for reducing participating enterprises' costs. Secondly, this paper used the GARCH model as a pre-training model of GRU model, and built a rolling training and testing framework. In this framework, the sample information can be updated and dynamically adjusted in time. Then we used recursion methods to evaluate and analyze the importance of input variables.

# 3 Variable selection

The proper input variables selection for the hybrid deep learning model will increase the accuracy of the prediction. In the following part, the influencing factors of the price are analyzed based on previous scholars' researches. This study will determine the primary input variables based on these analyses.

## 3.1 Input factors

Based on the previous studies, six categories of factors affect the price of carbon emission rights. They are carbon trading market, macroeconomics, energy market, futures market, industrial index, and air quality.

a. **Carbon trading market factors**

Due to the related factors such as different carbon emission reduction policies and economic development levels in various carbon pilot areas within the Chinese carbon trading market, the development degree of carbon finance shows differentiation. The price fluctuations of carbon trading in each pilot market vary significantly in space, and the price of carbon trading fluctuates sharply in time. For example, the carbon price in Shanghai and Shenzhen fluctuates fiercely while the carbon price in Guangdong is relatively stable. Meanwhile, the prices of different pilot markets may have a specific impact on each other. The carbon trading markets in these three mentioned regions have been established for the longest time and are the most representative. Therefore, considering the carbon trading market factors, the historical price variables of the market [1] were selected as the input variable of the model.

b. **Macroeconomy**

The fluctuation of economic activities is a determining factor to influence the price fluctuation of carbon emission rights. Theoretically, with the increasing prosperity of the macroeconomic environment, the carbon emissions of the industrial sector and the demand for carbon emission rights will grow more rapidly, accordingly leading to the rise of the price of emission rights. In economics, stock market prices and foreign exchange prices are always selected to investigate the development of the macroeconomy. Thus, this paper selected the Shanghai and Shenzhen 300 Index, Standard Poor's 500 Index, which can reflect the country's economic conditions and overseas. The foreign exchange rate of the domestic currency was also considered as a macroeconomic input variable [3].

c. **Energy market**

The industrial sectors usually present a high demand for carbon emission rights, especially the energy, steel, power generation, and other industries with high energy consumption. With the growth of industrial output, related carbon dioxide emissions will increase. As a result, more companies will purchase carbon emission rights to offset their emissions, causing the corresponding increase in the trading price of carbon emission rights. For the availability of data, this paper selected the retail price of LNG in Shanghai, the national gasoline spot benchmark price, the Australian BJ thermal coal spot price, the British Brent Crude oil prices, and the Wilder Global New Energy Index [4] as input variables in terms of the energy element.

d. **Futures market**

Commonly, the price of the energy futures market represents the vane of spot prices, and futures possesses the function of pricing. Thereby, domestic natural gas futures, carbon emission futures,

EU CER settlement prices, and carbon allowance EUA settlement prices were chosen as input variables [2].

**e. Industrial index**

Similar to the transmission mechanism of the macroeconomy, the rise in the industrial index indicates the prosperity of the industry, continuously leading to the upward trend in the carbon emissions within the industrial sector, the requirement for carbon emission rights, and carbon emission trading price. In this sense, this paper selected the SSE Energy Industry Index, the SSE Industrial Index, the SSE Natural Resources Index, the Domestic Mining Industry Index, the Shenzhen Stock Exchange Energy Index, and the Domestic Manufacturing Index as input variables [3].

**f. Air quality**

There are five major pollution standards for air quality in evaluating: ground-level ozone, particulate pollution, carbon monoxide, sulfur dioxide, and nitrogen dioxide. Due to the increase in production activities, the air pollutants continue to mount up, suggesting the growing tendency to carbon dioxide emission during the energy consumption process and the demand for carbon emission allowances. Furthermore, the PM2.5 index in Shanghai was considered a critical environmental factor. It directly impacts the price in the Shanghai carbon trading market. Therefore, this factor was selected as an input variable in this study [5].

In summary, this research selected 23 variables in six vital aspects: carbon trading market, macroeconomy, energy market, futures market, industrial index, and air quality. To study the latest carbon trading situation, we set the sample period from November 28th, 2016, to November 20th, 2020. Concerning the consistency and continuity of the data between the variables, the no-trade date data of the carbon trading market has been eliminated, and the missing values of the remaining variables were filled in with the average matter of the two trading days before and after the deleted figures, resulting in a total of 605 days of data. Table 1 shows the variables and their descriptions. All the data were collected from Wind and carbon emissions trading websites (http://www.tanpaifang.com/index.html).

**Table 1:** Variables and descriptions.

| Group | Variable | Description |
|---|---|---|
| Carbon Trading Market | SHEA | Shanghai carbon emission quota: daily closing price (yuan/ton) |
| | SZA | Shenzhen carbon emission quota: daily closing price (yuan/ton) |
| | GDEA | Guangdong carbon emission quota: daily closing price (yuan/ton) |
| Macro Economy | HS300 | Shanghai and Shenzhen 300 index |
| | BP500 | Standard & Poor 500 Index |
| | OY | The central parity rate of euro against yuan |
| | MY | The central parity rate between dollar and yuan |
| Energy Market | CCI500 | Thermal coal price index (tax included) (yuan/ton) |

| | YHQ | Shanghai Market Price of LPG (Imported Gas) (Yuan/ton) |
| --- | --- | --- |
| | QY | Domestic spot market price of gasoline (yuan/ton) |
| | FOB | UK Brent crude oil price (USD/BBL) |
| | WIRED | Weld Global New Energy Index |
| Futures Market | TRQQH | Natural gas futures |
| | TPFQH | Carbon futures |
| | CER | Certified Emission-Reduction Continuous Futures Contract Price (EUR/Ton of CO2) |
| | EUA | EU Emission Allowance Continuous Futures Contract Price (EUR/Ton of CO2) |
| Industrial Index | SZNY | Shanghai energy sector index |
| | SZGY | Shanghai Industrial Index |
| | SZZR | Shanghai natural resources index |
| | CKY | Domestic mining index |
| | SSZNY | Shenzhen Energy Index |
| | ZZY | Domestic manufacturing index |
| Air Quality | PM | Daily mean of PM2.5 in China (microgram/ton of carbon dioxide) |

## 3.2 Descriptive statistics and testing

According to the descriptive statistics on the data, the average transaction price of Shanghai Emissions Allowances (SHEA) is close to 36.92 yuan, with an approximate standard deviation of 5%. The normality test shows that most variables are significantly different from the normal distribution, and the results of an ADF test (significance level of 5%) after the first-order difference of data indicate that all the time series data are stable. In addition, the ARCH-LM test is a heteroscedasticity test that is used to identify whether the time series has the ARCH effect. The 12-order ARCH-LM test result of the Shanghai carbon market's transaction price has a P value of less than 0.001, rejecting the null hypothesis, which represents the existence of the ARCH effect. Moreover, there are apparent distinctions among the 23 input variables in terms of the statistical properties. These complex characteristics reveal that it is challenging to process time-series forecasts with the traditional statistical method. Therefore, an effective combination of traditional statistical methods (such as GARCH) and deep learning methods (such as GRU, LSTM) can play to their respective advantages and improve the model's predictive ability. Table 2 shows the details of the data set.

**Table2:** Descriptive statistics of the variables.

| Variable | Mean | Maximum | Minimum | Std. Dev. | Jarque-Bera | ADF |
| --- | --- | --- | --- | --- | --- | --- |
| SHEA | 36.92379 | 49.5 | 17.22 | 5.350171 | 140.9295 | -16.42103 |
| SZA | 23.88701 | 42.27 | 3.3 | 8.986592 | 32.71232 | -12.92713 |
| GDEA | 18.78742 | 30.84 | 1.27 | 6.256303 | 53.67231 | -16.60822 |
| CER | 0.242213 | 0.35 | 0.16 | 0.043531 | 36.32629 | -30.01796 |
| EUA | 16.49534 | 30.44 | 4.3 | 8.75704 | 65.73757 | -24.30287 |
| HS300 | 3835.727 | 4981.35 | 3035.874 | 435.0513 | 61.40266 | -22.58138 |
| BP500 | 2777.784 | 3626.91 | 2191.08 | 339.7764 | 21.19443 | -15.4346 |
| CCI500 | 545.8037 | 691 | 403 | 50.25032 | 13.39984 | -9.933255 |
| YHQ | 4489.95 | 5950 | 3150 | 512.3757 | 0.143388 | -21.21986 |
| QY | 6802.395 | 9445 | 5249 | 870.8075 | 32.88483 | -12.33075 |
| FOB | 57.50675 | 85.16 | 9.12 | 13.61086 | 36.02607 | -24.01927 |
| WIRED | 174.7168 | 253.82 | 144.75 | 17.48455 | 905.628 | -20.42274 |
| OY | 7.716362 | 8.2882 | 7.227 | 0.22797 | 4.004422 | -24.24129 |
| MY | 6.791517 | 7.1316 | 6.2764 | 0.214367 | 58.46737 | -23.3674 |
| PM | 35.51248 | 191 | 8 | 24.41435 | 1345.267 | -16.5554 |
| SZNY | 1404.261 | 1868.42 | 967.48 | 228.7053 | 39.86322 | -15.947 |
| SZGY | 2485.491 | 2935.13 | 1958.54 | 210.6122 | 4.911798 | -23.50572 |
| SZZR | 1865.917 | 2491.29 | 1415.72 | 242.4956 | 29.43513 | -25.09943 |
| CKY | 2013.172 | 2852.71 | 1467.93 | 341.1187 | 42.93306 | -25.30999 |
| SSZNY | 3250.209 | 4745.77 | 2157.29 | 762.817 | 61.63266 | -24.58108 |
| ZZY | 2035.428 | 2841.71 | 1362.45 | 313.1833 | 36.84538 | -23.35864 |
| TRQQH | 2.727719 | 4.715 | 1.482 | 0.560973 | 7.876429 | -24.56591 |
| TPFQH | 15.80171 | 29.75 | 4.3 | 8.562725 | 65.80915 | -25.29708 |

Note: The cut-off value of the Jarque-Bera test at 5% is 5.99. ADF is a stationarity test with a critical value of -2.86 at 5%. The ARCH (12) statistic corresponds to the ARCH-LM test with a lag of 12 order, which is represented here by the P-value.

Figures 1 and 2 illustrate the historical transaction prices and fluctuations of the Shanghai carbon trading market. From November 28th, 2016, to November 20th, 2020, the Shanghai Carbon Allowance transaction price(SHEA-2013) displayed an overall upward trend with four prominent price peaks, successively taking place in the first half of 2017, July 2018, the second half of 2019 and May 2020. During the sample period, the overall price fluctuates wildly. As the volatility rate is as high as 0.35 or above, a certain degree of volatility clustering could be observed, with the highest price at around 49 yuan per ton while the lowest price at less than 20 yuan per ton. Moreover, the overall difference is more than two times, thus indicating an obvious price abnormality. At the end of each year, the carbon trading price fell. It is because the current carbon emission rights are time-limited. Companies will clear the remaining carbon emission rights at the end of the year. Due to the quota carry-over policy, the price of SHEA-2013 surged at the beginning of 2017. The price further fluctuated sideways during the performance period. After the performance, the price fell back and returned to the upward trend at the end of the year. The transaction was generally active, with the growth of both volume and price. Before the performance period, the lowest transaction price was 24.75 yuan per ton, and the highest transaction price was 42 yuan per ton, fluctuating between 35-39 yuan/ton. Although carbon prices in Shanghai have fluctuated to a certain extent since 2018, an overall rising trend has been witnessed, with carbon prices stabilizing between 30 yuan per ton to 50 yuan per ton.

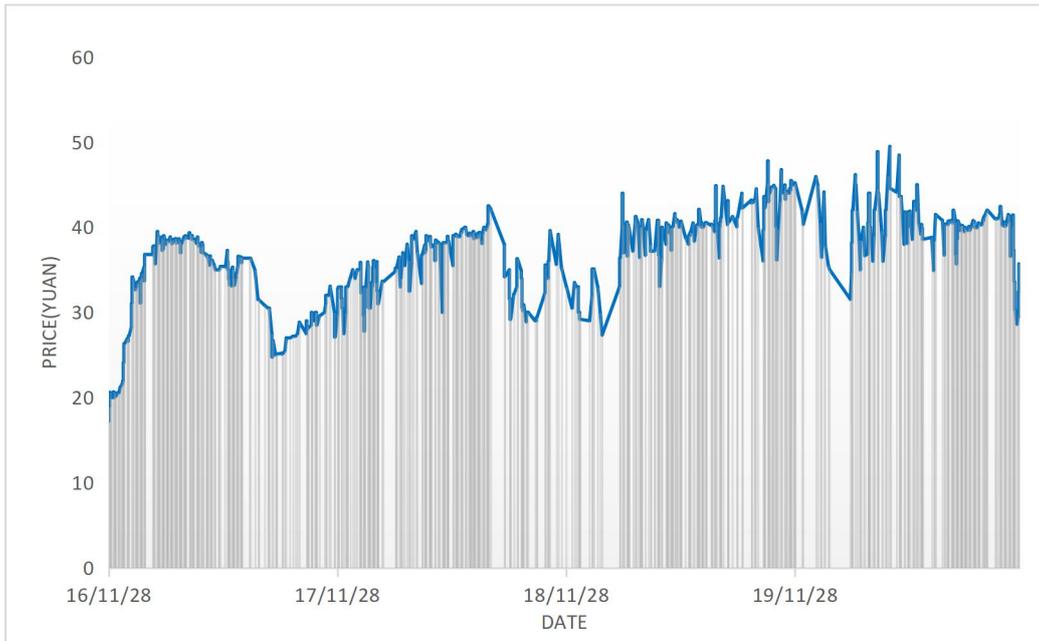

**Figure 1:** Transaction price of carbon trading in Shanghai.

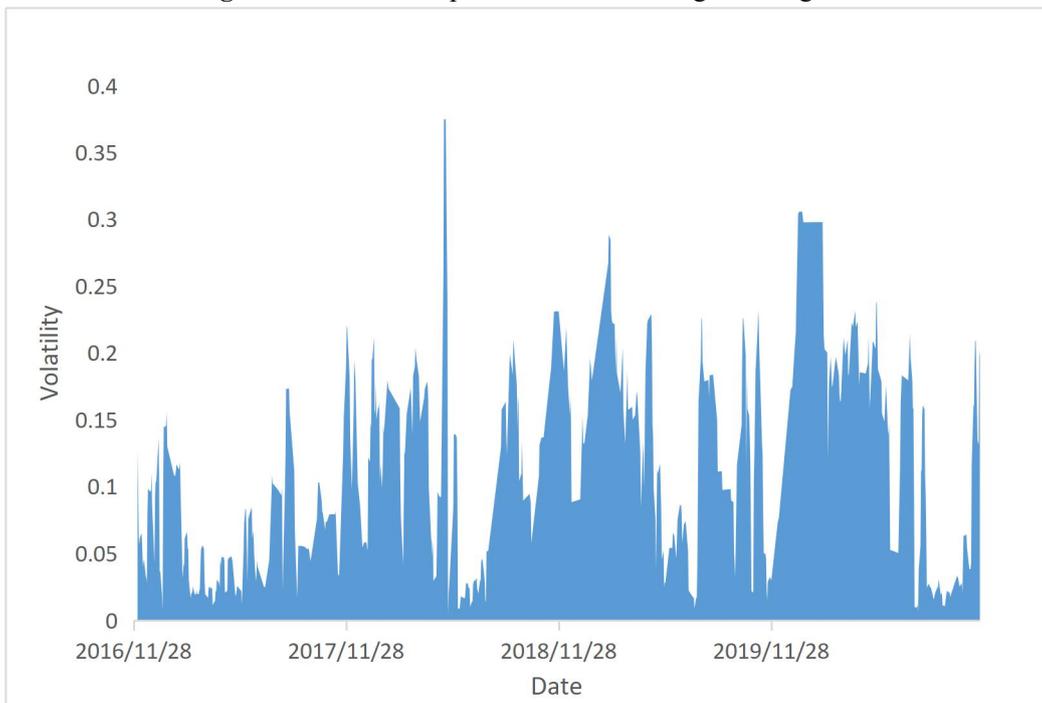

**Figure 2:** Volatility of carbon trading price in Shanghai.

# 4 Strategy

After completing the selection of variables and the description of the data, a hybrid deep learning model was established to construct a carbon emission trading strategy in this section. As shown in figure 3, first, we use the recursive feature elimination method [23] and obtained the variables that have significant contributions to the prediction through the progressive measurement of importance. Subsequently, we constructed a hybrid deep learning model (combining GARCH and GRU) and used it to predict the price of carbon emission rights. By comparing it with GARCH,

GRU, LSTM, moving average method, the reliability of the carbon trading price prediction model was also evaluated. Finally, in this section, based on the iceberg order theory [27], the timing signals of each model were generated to construct a scientific carbon emission rights purchasing strategy for carbon-deficient companies to reduce purchasing costs. The results of the five methods were compared and evaluated. The software used in this paper is Python 3.8, R 4.0.3, and the research flowchart is as follows.

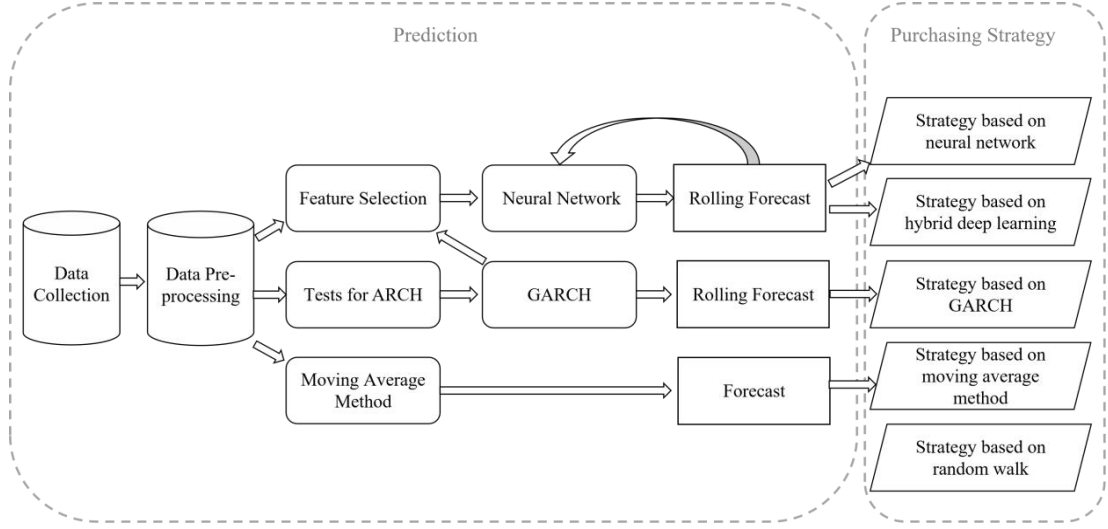

**Figure 3:** Research flowchart.

## 4.1 Methods
### 4.1.1 Hybrid deep learning

The hybrid deep learning model in this section refers to the combination of GARCH and GRU (GARCH-GRU), which means the price of carbon emission rights predicted by the GARCH model for the day is used as the input variable of GRU. This combination will make full use of the structural advantages and parameter estimation advantages of the GARCH model to achieve rapid pre-training of the hybrid model. It provides adequate information and essential variables for the hybrid deep learning model. For evaluating the effectiveness of this combination, the following study will replace GRU with LSTM to construct GARCH-LSTM, and compare the predictive capabilities and strategy implementation effects of the two. The following is a detailed introduction to single models.

*GARCH model*

The GARCH model was first proposed in 1986 [7]. The GARCH model is a regression model specifically for financial data. It can model the variance of errors and is especially suitable for volatility analysis and forecasting. This model assumes that the conditional variance is a function of the lag residual squared. The specification for the GARCH *(p, q)* is defined as:

$$r_t = c_1 + \sum_{i=1}^{R} \phi_i r_{t-i} + \sum_{j=1}^{M} \phi_j \varepsilon_{t-j} + \varepsilon_t \qquad (1)$$

$$\varepsilon_t = \mu_t \sqrt{h_t} \qquad (2)$$

$$h_t = k + \sum_{i=1}^{q} G_i h_{t-i} + \sum_{i=1}^{p} A_i \varepsilon_{t-i}^2 \qquad (3)$$

Among them, $h_t$ is the conditional variance, $\mu_t$ is an independent and identically distributed random variable, $h_t$ and $\mu_t$ are independent of each other, and $\mu_t$ is normally distributed.

*LSTM model*

The Long Short Term Memory Network (LSTM) is a particular Recurrent Neural Network (RNN) [13]. Its core concept lies in the cell state and the structure of the *gate*（Figure 4）. The cell state plays the role of the information transmission path, which allows information to be continuously transmitted in the sequence. It can be regarded as a *memory* on the Internet. The LSTM network is composed of interconnected unit blocks. Each unit block contains three types of gates: input gates, output gates, and forget gates, which implement writing, reading, and resetting the memory cells, respectively. The activation function controls these gates. The value of the activation function is 0 for a complete inhibition and the value 1 for a complete activation.

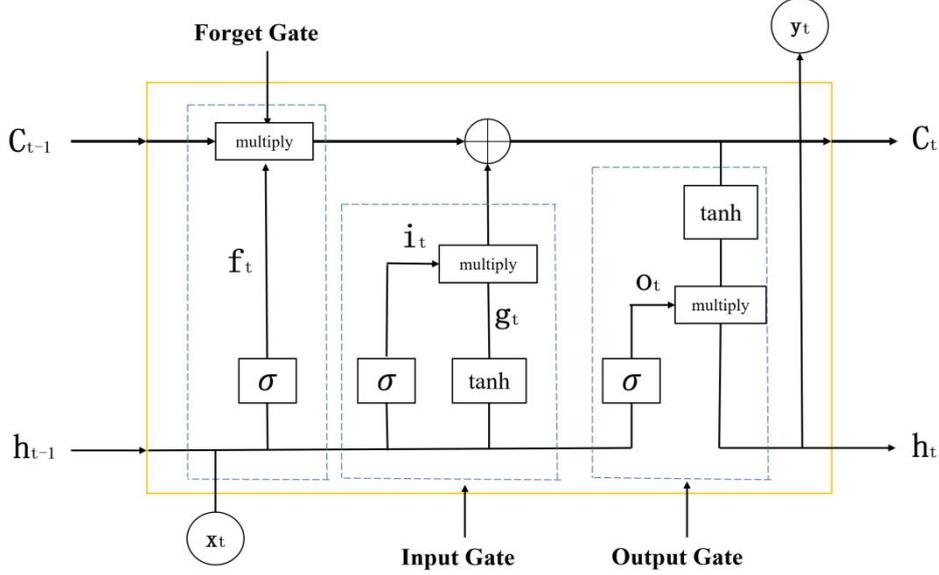

**Figure 4:** An illustrative graph of the LSTM unit.

Among them, $h_t$ is a short-term state, $C_t$ is a long-term state, $f_t$ represents the degree of forgetting the last information $C_{t-1}$, and $i_t$ represents the degree of retention of newly added information. This step realizes the removal of old information and the addition of new information. The equations for an LSTM unit are demonstrated in (4)-(9):

$$i_t = \sigma(W_{xi}^T \cdot x_t + W_{hi}^T \cdot h_{t-1} + b_i) \tag{4}$$

$$f_t = \sigma(W_{xf}^T \cdot x_t + W_{hf}^T \cdot h_{t-1} + b_f) \tag{5}$$

$$o_t = \sigma(W_{xo}^T \cdot x_t + W_{ho}^T \cdot h_{t-1} + b_o) \tag{6}$$

$$g_t = \tanh(W_{xg}^T \cdot x_t + W_{hg}^T \cdot h_{t-1} + b_g) \tag{7}$$

$$c_t = f_t \otimes c_{t-1} + i_t \otimes g_t \tag{8}$$

$$y_t = h_t = o_t \otimes \tanh(c_t) \tag{9}$$

In the formula, $W_{xi}$, $W_{xf}$, $W_{xo}$ and $W_{xg}$ are weight matrixes in the four layers connected to the input vector $x_t$. $W_{hi}$, $W_{hf}$, $W_{ho}$ and $W_{hg}$ are weight matrixes in the four layers attached to the short-term state $h_{t-1}$. $b_i$, $b_f$, $b_o$ and $b_g$ are the deviation items for each of the four layers.

*GRU model*

Gate Recurrent Unit (GRU) is a type of recurrent neural network (RNN) [28]. The same as LSTM, it is also proposed to solve long-term memory and gradients in backpropagation. It can effectively capture long sequences. The semantic association between the two can alleviate the phenomenon of gradient disappearance or explosion. It is more straightforward in calculation than LSTM, which can significantly improve the calculation efficiency.

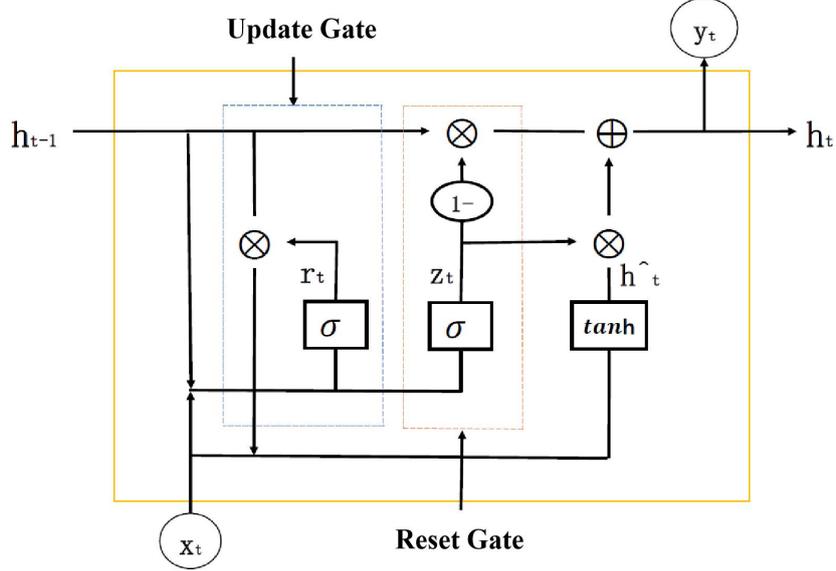

**Figure 5:** An illustrative graph of the GRU unit.

GRU optimizes the three *gates* of LSTM, and only retains two, namely the reset gate and the update gate (Figure 5). The reset gate determines how to combine the new input information with the previous memory. The update gate defines the amount of the previous memory saved to the current time step. If the reset gate is set to 1, and the update gate is set to 0, the result is a standard RNN model. The equations for a GRU unit are demonstrated in (10)-(13):

$$r_t = \sigma(W_{ir}x_t + b_{ir} + W_{hr}h_{(t-1)} + b_{hr}) \qquad (10)$$

$$z_t = \sigma(W_{iz}x_t + b_{iz} + W_{hz}h_{(t-1)} + b_{hz}) \qquad (11)$$

$$n_t = \tanh(W_{in}x_t + b_{in} + r_t * (W_{hn}h_{(t-1)} + b_{hn})) \qquad (12)$$

$$h_t = (1-z_t)*n_t + z_t * h_{(t-1)} \qquad (13)$$

**4.1.2 Forecast evaluation**

To evaluate and compare the performance of prediction models, we use five commonly used prediction error metrics, including MAE, MSE, MAPE, MSPE, LL. Their definitions are as follows:

Table 3: Forecast evaluation indicators.

| Indicator | Name | Formula |
|---|---|---|
| MAE | Mean Absolute Error | $MAE = \frac{1}{N}\sum_{i=1}^{N}|PV_t - RV_t|$ |
| MSE | Mean Squared Error | $MSE = \frac{1}{N}\sum_{i=1}^{N}(PV_t - RV_t)^2$ |
| MAPE | Mean Absolute Percentage Error | $MAPE = \frac{1}{N}\sum_{i=1}^{N}|1 - PV_t/RV_t|$ |
| MSPE | Mean Squared Percentage Error | $MSPE = \frac{1}{N}\sum_{i=1}^{N}(1 - \frac{PV_t}{RV_t})^2$ |
| LL | Logarithmic Loss | $LL = \frac{1}{N}\sum_{i=1}^{N}(\ln(PV_t) - \ln(RV_t))^2$ |

Among them, $PV_t$ and $RV_t$ are the predicted value and the real value of the carbon allowance transaction price respectively. N is the number of predictions. The smaller the metric value of the prediction result is, the higher the prediction accuracy is.

In Table 3, MAE, MSE, MAPE, MSPE, and LL belong to the family of symmetric loss functions, in the sense that they equally penalize over and under predictions. The most widely used indicator, MSE, is proposed by Bollerslev et al. [24]. This error measurement method is more sensitive to severe individual prediction errors. While MAE measures the absolute average error, and this method is less susceptible to extreme individual error predictions than MSE. MAPE and MSPE were proposed by Makridakis [25], adding measures of percentage errors. In addition, the log loss (LL) introduced by Pagan and Schwert [26], which penalizes under-prediction, is greater than the penalty for over-prediction.

**4.1.3 Feature generation and model construction**

Combining GARCH and GRU, the GARCH-GRU model is built. We use the GARCH model to predict the daily price of carbon emission rights on a rolling basis with 200-day historical prices, and use it as the input feature of the GRU model. To be exact, we use the historical prices of i$^{th}$ to i+199$^{th}$ days to establish GARCH model, which can predict the price of i+200$^{th}$ day. Then we put this GARCH result and the other variables of i+199$^{th}$ day into the corresponding unit of GRU model as input variables, to conduct the final predictions. Before the model enters the implementation set, we use *GridSearchCV* method to tune three parameters for obtaining the optimal parameters: dropout is 0.2, epoch is 150, and learning rate is 0.01. In addition, the *recursive feature elimination* method [23] is used to select feature screening on 23 input variables. Four variables (FOB, WIRED, SZGY, and PM) are eliminated, which have the weakest contributions to the prediction. We replace the historical closing price of SHEA with the day price predicted by GARCH, then the new input variable of the hybrid model is obtained.

To improve the prediction accuracy, we used the rolling dynamic prediction method. The usual static prediction method divides the data set into a training set and a test set. The training set generates a model, and the model evaluation is performed in the test set. Different from the past, the data construction arrangement under our method is as follows. First, the input variables are matched with the output variables. As shown in Figure 6, the output variable of the carbon emission price $n_1+1$ corresponds to 19 variable data from the 1$^{st}$ to $n_1^{th}$ trading day, where $n_1$ is called the sliding time window. These variables together constitute one piece of data. The training set (such as $TR_1$) required for one rollover consists of several pieces of such data. We will get this rolling model through training and use it to predict the price of carbon emission rights on the next trading day. In other words, each rolling forecast is only one day backward. After completing the first rolling, the data of all training sets will be updated by one trading day back. The new data is used to generate a new prediction model, which can continue to predict the price of carbon emission rights on the next trading day. Such dynamic rolling forecast will continue to loop until the data of the last trading day. It is worth noting here that this paper sets a segmentation point for the data set. Before the segmentation point, it is called the tuning set. In this part, model tuning is allowed, and the second half is the implementation set. This part can only be used for rolling training and prediction to focus on the effectiveness of the objective evaluation model in the implementation. The description of each parameter in the data will be further given below.

To be exact, We first set the sliding time window to $n_1$ days, and predict the number of steps to be 1 day. The $n_1$ can be 5, 10, and 20. The number of days in once rolling is $n$, and the data is divided into the training set $TR_n$ and the test set $TE_n$. The total sample size $N$ of data is 605, which means, the total train and prediction will roll $N-n$ times. This study set the first 70% rolling data as the parameter set, and the rest 30% rolling data as the implementation set. For example, $n_1$ is 5, $n$ is 60, then the number of parameter sets is 70% of 545 results, which is equal to 381. The process is shown in Figure 6.

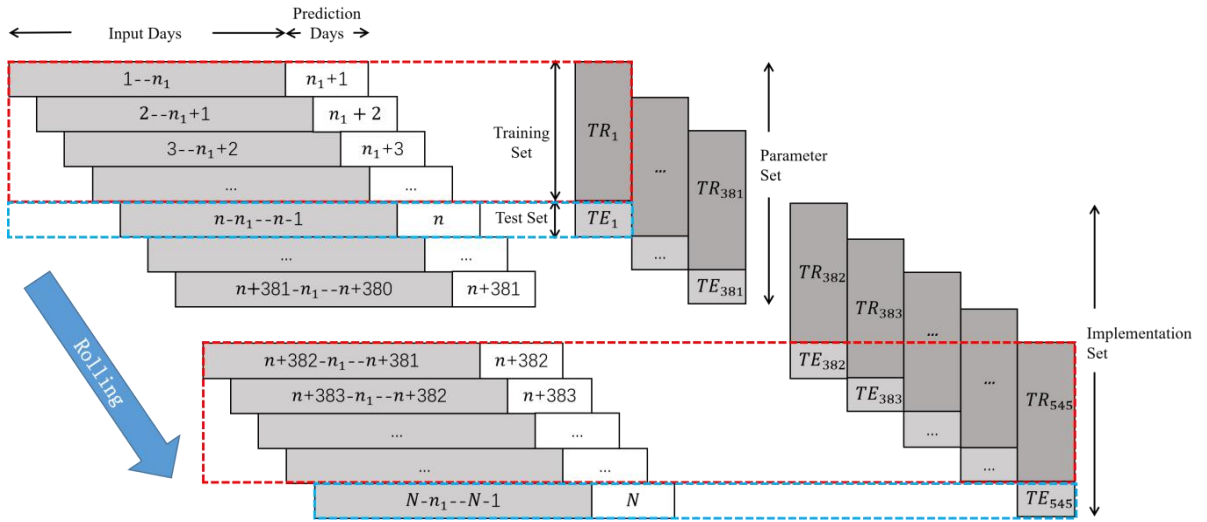

**Figure 6:** Schematic of rolling forecast.

Note: The first row on the left is the first rolling, which means the model is trained in the data between 1st day to $n_1$th day for predicting the price at the $n_1+1$th day, $n$ is the sample size in one rolling. $TR_n$ and $TE_n$ are the training and test sets in the $n$th rolling. $N$ is the total sample size of the whole data set.

### 4.1.4 Comparison of model results

We select five models (GARCH, Moving Average, GRU, LSTM, and GARCH-LSTM) to compare with GARCH-GRU, and make predictions under sliding time windows with 5, 10, and 20 days. The results are shown in Figures 7 to 9.

From Table 4, the following results can be observed. First of all, the error of GARCH is higher than the moving average method and better than some GRU and LSTM models. Secondly, inputting GARCH predicted values as features into the deep learning model can profoundly improve the model's predictive ability. Compared with the GARCH model or the deep learning model, better prediction results can be obtained from the hybrid deep learning model. Among the six models in the table, the GARCH-GRU model with a 5-day sliding time window has the highest forecast progress. Its MAE is 1.1713, MSE is 2.5358, MAPE is 2.8998, MSPE is $1.62 \times 10^{-3}$, and LL is $1.58 \times 10^{-3}$.

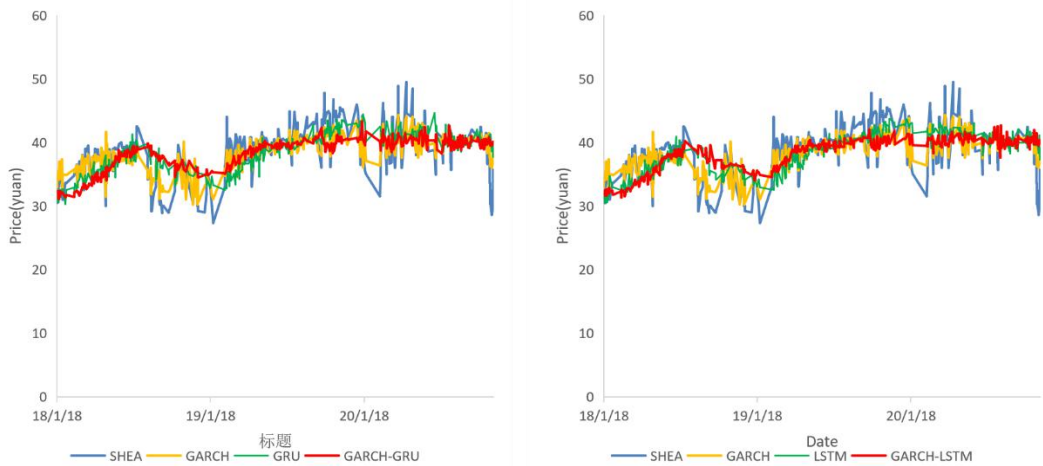

**Figure 7:** Predicted and realized price of window 5.

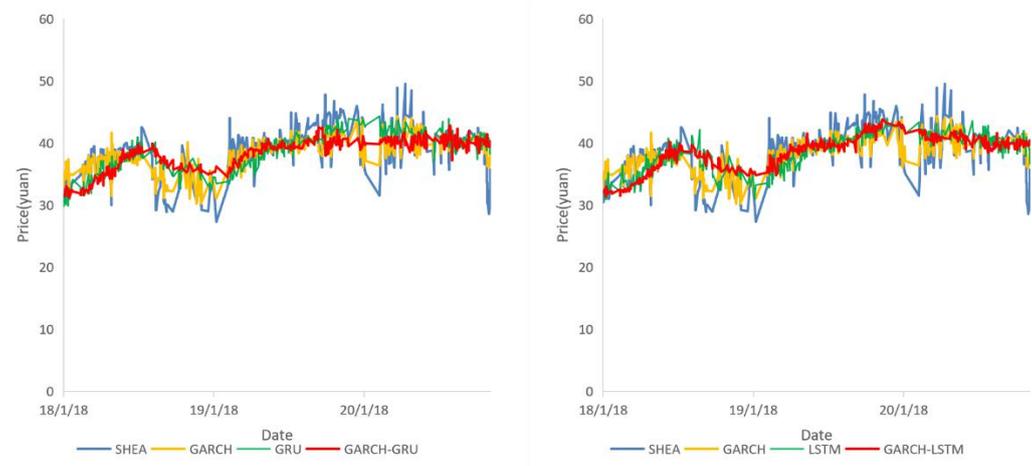

**Figure 8:** Predicted and realized price of window 10.

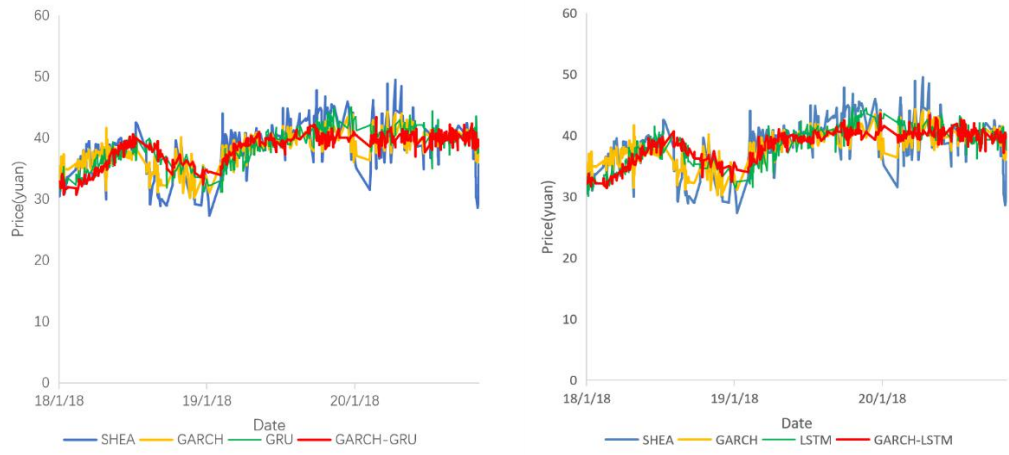

**Figure 9:** Predicted and realized price of window 20.

Table 4: Prediction performances.

| Indicator | | MAE | MSE | MAPE | MSPE | LL |
|---|---|---|---|---|---|---|
| Model | Window | | | | | |
| GARCH | 200 | 1.6211 | 4.4008 | 4.2627 | $3.08 \times 10^{-3}$ | $3.10 \times 10^{-3}$ |
| MA | 5 | 1.6310 | 5.8470 | 4.5170 | $4.49 \times 10^{-3}$ | $4.49 \times 10^{-3}$ |
| | 10 | 1.8250 | 6.8140 | 5.1344 | $5.53 \times 10^{-3}$ | $5.56 \times 10^{-3}$ |
| | 20 | 2.1946 | 9.4991 | 6.2433 | $8.37 \times 10^{-3}$ | $8.18 \times 10^{-3}$ |
| GRU | 5 | 2.4418 | 11.4040 | 5.9134 | $8.55 \times 10^{-3}$ | $7.44 \times 10^{-3}$ |
| | 10 | 2.3810 | 11.2371 | 5.7418 | $8.79 \times 10^{-3}$ | $7.43 \times 10^{-3}$ |
| | 20 | 2.5886 | 12.2961 | 6.3096 | $9.08 \times 10^{-3}$ | $7.98 \times 10^{-3}$ |
| LSTM | 5 | 2.4821 | 11.8352 | 6.0245 | $8.80 \times 10^{-3}$ | $7.72 \times 10^{-3}$ |
| | 10 | 2.5129 | 11.2841 | 6.1137 | $8.54 \times 10^{-3}$ | $7.43 \times 10^{-3}$ |
| | 20 | 2.5282 | 11.8016 | 6.1798 | $8{,}82 \times 10^{-3}$ | $7.76 \times 10^{-3}$ |
| GARCH-GRU | 5 | **1.1713** | **2.5358** | **2.8998** | $1.62 \times 10^{-3}$ | $\mathbf{1.58 \times 10^{-3}}$ |
| | 10 | 1.2338 | 2.7163 | 3.0544 | $1.75 \times 10^{-3}$ | $1.69 \times 10^{-3}$ |
| | 20 | 1.2626 | 2.9914 | 3.1351 | $1.89 \times 10^{-3}$ | $1.87 \times 10^{-3}$ |
| GARCH-LSTM | 5 | 1.2897 | 2.8744 | 3.2009 | $1.85 \times 10^{-3}$ | $1.80 \times 10^{-3}$ |
| | 10 | 1.4674 | 3.6709 | 3.5710 | $2.37 \times 10^{-3}$ | $2.24 \times 10^{-3}$ |
| | 20 | 1.4165 | 3.3507 | 3.5183 | $2.15 \times 10^{-3}$ | $2.08 \times 10^{-3}$ |

## 4.2 Local interpretation of variables

To explore the local interpretability of the input variables of the carbon emission rights pricing model further, we used the recursive elimination method to eliminate 19 features one by one. We calculated the adjustment of predictive ability and then evaluated the importance of the variables in the model prediction. In this paper, the error of the GARCH-GRU model with a sliding time window of 5 after excluding a single variable is shown in Table 5, and the ranking is shown in Table 6.

**Table 5:** Prediction performances when the single variable is deleted.

| Deletion of Variable | MAE | MSE | MAPE | MSPE | LL |
|---|---|---|---|---|---|
| **GSHEA** | **1.8454** | **6.6665** | **4.5167** | $\mathbf{5.11 \times 10^{-3}}$ | $\mathbf{4.49 \times 10^{-3}}$ |
| CKY | 1.5207 | 3.5863 | 3.7266 | $2.24 \times 10^{-3}$ | $2.18 \times 10^{-3}$ |
| TPFQH | 1.4814 | 3.6484 | 3.6246 | $2.30 \times 10^{-3}$ | $2.22 \times 10^{-3}$ |
| MY | 1.5055 | 3.5887 | 3.6848 | $2.25 \times 10^{-3}$ | $2.18 \times 10^{-3}$ |
| TRQQH | 1.513 | 3.5421 | 3.7065 | $2.21 \times 10^{-3}$ | $2.15 \times 10^{-3}$ |
| OY | 1.4788 | 3.4962 | 3.6175 | $2.19 \times 10^{-3}$ | $2.12 \times 10^{-3}$ |
| QY | 1.4745 | 3.435 | 3.5992 | $2.20 \times 10^{-3}$ | $2.10 \times 10^{-3}$ |
| GDEA | 1.4725 | 3.4741 | 3.6085 | $2.19 \times 10^{-3}$ | $2.11 \times 10^{-3}$ |
| ZZY | 1.4754 | 3.4287 | 3.6134 | $2.15 \times 10^{-3}$ | $2.09 \times 10^{-3}$ |
| EUA | 1.4752 | 3.3947 | 3.6108 | $2.16 \times 10^{-3}$ | $2.08 \times 10^{-3}$ |
| SSZNY | 1.4655 | 3.4175 | 3.5813 | $2.16 \times 10^{-3}$ | $2.08 \times 10^{-3}$ |
| CCI500 | 1.4282 | 3.3658 | 3.4762 | $2.18 \times 10^{-3}$ | $2.06 \times 10^{-3}$ |
| SZZR | 1.4343 | 3.3278 | 3.5103 | $2.09 \times 10^{-3}$ | $2.02 \times 10^{-3}$ |
| CER | 1.429 | 3.1788 | 3.4892 | $2.05 \times 10^{-3}$ | $1.96 \times 10^{-3}$ |
| YHQ | 1.4142 | 3.2489 | 3.4565 | $2.08 \times 10^{-3}$ | $1.99 \times 10^{-3}$ |
| BP500 | 1.4003 | 3.1565 | 3.4078 | $2.02 \times 10^{-3}$ | $1.93 \times 10^{-3}$ |
| SZNY | 1.4069 | 3.1297 | 3.4488 | $1.96 \times 10^{-3}$ | $1.90 \times 10^{-3}$ |
| SZA | 1.3921 | 3.169 | 3.4166 | $1.96 \times 10^{-3}$ | $1.92 \times 10^{-3}$ |
| HS300 | 1.3991 | 3.1304 | 3.4151 | $2.00 \times 10^{-3}$ | $1.91 \times 10^{-3}$ |

As shown in Table 5, if we eliminate one of the 19 variables, we can figure out new results of the GARCH-GRU model with a 5-day window. When the variable the GARCH rolling forecast value of the Shanghai carbon trading price (GSHEA) is eliminated, the MAE of the model extends to 1.8454, the MSE reaches 6.6665, and the MAPE, MSPE, LL are respectively 4.5167, $5.11 \times 10^{-3}$, and $4.49 \times 10^{-3}$. Obviously, the variable that has the most significant impact on the price of carbon emission rights in the Shanghai pilot market is the GSHEA (Table 6). It is followed by the domestic mining index (CKY), the carbon futures (TPFQH), and the central parity rate between dollar and yuan (MY). The variable GSHEA ranks No. 1, which means that the inclusion of the GARCH prediction value of the Shanghai carbon trading price has improved the accuracy of the model significantly. The domestic mining index is an industrial index, which is also closely associated with carbon emission because mining machines can let out large amounts of carbon emission. The future price always predict the spot price, so the carbon futures price is related to real carbon emission price. In addition, the central parity rate between dollar and yuan is a macroeconomic factor. We can conclude that domestic and foreign macroeconomics have an impact on the pricing of carbon emission rights. Furthermore, preliminary conclusions can be drawn from the 19 indicators selected: the carbon trading market, the macroeconomy, the futures market, the energy market, and the industrial index, all impact the formation of carbon emission rights prices. Besides, air quality has no significant effect on the cost of carbon emission rights for the time being.

Table 6: Ranking of the importance of all variables.

| Ranking | MAE | MSE | MAPE | MSPE | LL | Average Ranking |
|---|---|---|---|---|---|---|
| **GSHEA** | **1** | **1** | **1** | **1** | **1** | **1** |
| CKY | 2 | 4 | 2 | 4 | 4 | 2 |
| TPFQH | 5 | 2 | 5 | 2 | 2 | 3 |
| MY | 4 | 3 | 4 | 3 | 3 | 3 |
| TRQQH | 3 | 5 | 3 | 5 | 5 | 5 |
| OY | 6 | 6 | 6 | 7 | 6 | 6 |
| QY | 9 | 8 | 10 | 6 | 8 | 7 |
| GDEA | 10 | 7 | 9 | 8 | 7 | 8 |
| ZZY | 7 | 9 | 7 | 12 | 9 | 9 |
| EUA | 8 | 11 | 8 | 11 | 10 | 10 |
| SSZNY | 11 | 10 | 11 | 10 | 11 | 11 |
| CCI500 | 14 | 12 | 14 | 9 | 12 | 12 |
| SZZR | 12 | 13 | 12 | 13 | 13 | 13 |
| CER | 13 | 15 | 13 | 15 | 15 | 14 |
| YHQ | 15 | 14 | 15 | 14 | 14 | 15 |
| BP500 | 17 | 17 | 19 | 16 | 16 | 16 |
| SZNY | 16 | 19 | 16 | 18 | 19 | 16 |
| SZA | 19 | 16 | 17 | 19 | 17 | 16 |
| HS300 | 18 | 18 | 18 | 17 | 18 | 16 |

## 4.3 Timing signal generation and strategy evaluation

The rolling prediction model has been used to predict the price of carbon emission rights. We would further construct the carbon emission rights purchasing strategy based on the prediction and Iceberg Order theory. Iceberg Order Theory [27] originally refers to quantitative trading. To avoid the impact of the market and hide their trading motivation, traders often divide the whole order into small batches to buy or sell. These orders are called *iceberg order*s. This interesting name implies that the most significant part of an iceberg floats is underwater and invisible. Only less than one-ninth of the ice is out of the water. For obtaining a better purchasing price, we innovatively applied this theory for carbon trading and provided a scientific and reasonable cost reduction purchasing strategy.

Specifically, for example, the Shanghai carbon trading market encourages companies to reduce carbon dioxide emissions voluntarily. Depending on the historical emission method and the baseline method, the emission quotas are allocated to pilot enterprises free of charge (generally issued annually). A risk control mechanism is established in the market, that can restrict the number of quota holders and ensure the smooth running of the Shanghai carbon market. Therefore, once enterprises have to emit more carbon dioxide due to production growth, technological reasons, or their technology emission reduction effect is insufficient, they need an extra-budgetary carbon quota. Generating timing signals and constructing a carbon emission right purchasing strategy to reduce enterprises' purchasing costs are the crucial problems to be solved in this paper.

In summary, we used the prediction results of the above six models in the sliding time window of 5 days to generate timing signals. According to the Iceberg Order Theory, a large number of carbon quotas are divided into small payments. When the *buy* signal appears, orders are placed in equal batches to complete the purchase of the shortage. The rules of timing signal generation are given below.

**4.3.1 timing signal generation**

Before carrying out the trading strategy, there are some assumptions added: We assume that the liquidity of the company is sufficient and there is no limit on the amounts and time of orders in the market. Timing signal (y) is a classification vector containing only two common trading signals: *buy* and *hold*. The timing signal is calculated by the forecast price based on a hybrid deep learning prediction model ($SHEA_{t+1}^{p}$). The calculation is as follows:

$$\Delta = \frac{SHEA_{t+1}^{P}}{SHEA_{t}} - 1 \qquad (14)$$

$$y_t = \begin{cases} 1 \cdots\cdots\cdots\cdots\cdots\cdots\cdots\cdots\cdots \Delta \geq 0.02 \\ 0 \cdots\cdots\cdots\cdots\cdots\cdots\cdots\cdots\cdots \Delta < 0.02 \end{cases} \qquad (15)$$

Among them, $SHEA_{t+1}^{p}$ represents the rolling forecast price of day t+1, $SHEA_{t}$ represents the rolling forecast price of day t. The threshold is set to 0.02. If $\triangle$ exceeds the threshold, the timing signal $y_t$ is gained as 1 and 0, respectively, which means buying and holding at time t in turn.

**4.3.2 strategy effectiveness evaluation**

To further evaluate the purchase strategy generated by GARCH-GRU, we introduced the purchasing strategy under random timing signal and analyze the purchase cost together with the purchase strategies of the other five models. Since carbon quotas are generally issued annually, we selected the data of nearly one year: November 4th, 2019, to November 5th, 2020, to simulate five kinds of strategies. Given that the annual carbon quota of enterprise A lacks 20000 tons, and every transaction will buy 1000 tons of carbon emission rights, so a total of 20 purchase transactions will be conducted. If the purchasing strategy has reached 20 purchase times, the transaction will be stopped automatically. Finally, the purchase cost is obtained by substituting the closing price of carbon emission right day in this year. The specific cost is shown in Table 7.

**Table 7:** Costs of 5 kinds of strategies.

| Purchasing Strategy | Total Cost (yuan) | Reduction Ratio | Relative Quantile of Random Strategy Cost |
|---|---|---|---|
| GARCH | 801710 | 2.28% | 5.4% |

| | | | | |
|---|---|---|---|---|
| Deep-Learning | GRU | 802320 | 2.20% | 5.9% |
| | LSTM | 804970 | 1.88% | 9.5% |
| Hybrid Deep-Learning | GARCH-GRU | 789710 | 3.74% | 0.7% |
| | GARCH-LSTM | 798220 | 2.70% | 3.2% |
| Moving Average | | 800940 | 2.37% | 4.8% |
| Mean Value of Random Strategy（1000 trials） | | 820396 | | |

Note: The random purchasing strategy is the average cost after 1000 simulations of 20 days randomly selected within one year.

From the simulation results, the purchase cost of the strategy based on the GARCH-GRU model is the least, which is 789,710 yuan, and it can be executed at a low price almost every time. Compared with the random purchasing strategy, the cost of the process based on the GARCH-GRU model reduces 3.74%. Next to it is the purchase strategy based on the GARCH-LSTM model, the required purchase cost is 798,220 yuan, and the reduction rate of purchasing cost is 2.7%. Other designs also have about 2% cost reduction. In addition, only 7 of the 1000 times of the random purchase strategy are less than 789,710 yuan. It means the GARCH-GRU based purchase strategy is superior to the 99.3% random strategy. To sum up, the purchasing strategy based on the GARCH-GRU model can effectively choose the appropriate time to buy carbon emission rights, and help enterprises reduce costs.

## 5 Conclusion

In summary, the price risk within the Chinese carbon trading markets demonstrates two significant features in terms of space and time. Firstly, there are substantial spatial differences in the trading prices of each carbon trading pilot region, implying the spillover risk of price fluctuation. Secondly, the price of carbon trading presents sharp fluctuations and volatile prices over time, leading to an increasing degree of difficulty in investment. Therefore, this research took Shanghai as an example and constructed a GARCH-GRU model to predict the transaction price of carbon trading, and then generated the purchase signal to frame the purchasing strategy of carbon emission rights. The conclusions are as follows:

These input variables can be divided into six categories: carbon trading market, macro-economy, energy market, futures market, industrial index, and air quality. After a process of recursive feature elimination, this study extracts 19 variables from the original 23 variables. Model obtains the dominant variables that have significant contributions to the prediction followed by measures of importance. Then, a hybrid deep learning model (GARCH-GRU) is formed to predict the price of carbon emission rights. With the comparisons of feasible approaches such as GARCH, GRU, LSTM, and moving average, this paper verifies that the deep learning model with GARCH is better than the other five models. Based on the Iceberg Order Theory, this paper generates the timing signals of five models. It separately constructs batch purchasing strategies of carbon emission rights for carbon-deficient enterprises to reduce the purchase cost. By comparing and evaluating the results of the five methods, the cost of the strategy based on the GARCH-GRU model is the lowest, which is superior to 99.3% of the random strategy under 1000 simulation transactions. Although the random number in the algorithm will have a small impact on the value of the calculation result, it does not affect the main conclusion of this paper.

Due to the price fluctuation in the carbon trading market, it is rather complicated to arrange emission reduction activities. While the price surge will inhibit emission reduction activities, the price slump is not conducive for enterprises to set emission reduction investments in the long term. Therefore, the prediction of the fluctuation in carbon emission trading price and the proposal of purchasing strategy provide references for the practical operation of companies. This paper took the Shanghai pilot market as an example and proposed a purchasing strategy based on a hybrid deep learning model. It will help improve the companys' awareness of carbon emission reduction continuously so that enterprises will adjust their production and operation activities based on accurate market predictions and reduce production costs. Meanwhile, the environmental conditions can also be improved due to the reasonable arrangement of emission reduction activities, resulting in the ultimate achievement in economic and social benefits.

**Author Contributions**
Conceptualization, Y.F.; methodology, Y.F. and J.X.; software, J.X.; writing—original draft preparation, Y.F. and J.X.; writing—review and editing, Y.F. and J.X. All authors have read and agreed to the published version of the manuscript.

**Data Availability**
The data used to support this study can be obtained from the authors upon request (**1000497252@smail.shnu.edu.cn**).

**Conflicts of Interest**
The authors declare that there are no conflicts of interest regarding the publication of this paper.